\documentclass[a4paper,traditabstract]{aa} 
\usepackage{graphicx}
\usepackage{color}
\usepackage{txfonts}
\usepackage{url}
\usepackage{natbib}
\bibpunct{(}{)}{;}{a}{}{,}
\usepackage[utf8]{inputenc}
\usepackage{subfigure}
\usepackage{verbatim}
\usepackage{amssymb}
\begin{document}
   \title{On the asymmetry of velocity oscillation amplitude\\ in bipolar active regions}
   \author{F. Giannattasio$^{1}$, M. Stangalini$^{1,2}$, D. Del Moro$^{1}$, F. Berrilli$^{1}$}
   \institute{$^{1}$Dipartimento di Fisica, Università di Roma "Tor Vergata", Via della Ricerca Scientifica,1 00133 Rome, Italy\\
   $^{2}$Max Planck Institute for Solar System Research, Max-Planck-Str. 2 37191 Katlenburg-Lindau Germany\\
   \email{Fabio.Giannattasio@roma2.infn.it}}
             
   \date{}

 
  \abstract
   {The velocity field in the lower solar atmosphere undergoes strong interactions with magnetic fields. Many authors have pointed out that power is reduced by a factor between two and three within magnetic regions, depending on frequency, depth, the radius and the magnetic strength of the flux tube. Many mechanisms have been proposed to explain the observations. In this work, SDO dopplergrams and magnetograms of 12 bipolar active regions ($\beta$ARs) at a 45 second cadence, are used to investigate the relation between velocity fluctuations and magnetic fields. We show that there is an asymmetry within $\beta$ARs, with the velocity oscillation amplitude being more suppressed in the leading polarities compared to the trailing polarities. Also, the strongest magnetic fields do not completely suppress the five-minute oscillation amplitude, neither in the spot innermost umbrae.}

   \authorrunning{F. Giannattasio et al.}
   \titlerunning{On the asymmetry of velocity oscillation amplitude\\ in bipolar active regions}

\maketitle

%

\section{Introduction}
The study of the interaction between waves and magnetic fields plays an important role in the investigation of the mechanisms responsible for the solar corona heating.\\
It is well established that the amplitude of solar five-minute oscillations in the photosphere is reduced by a factor of two to three within ARs \citep{1962ApJ...135..474L, 1981SoPh...69..233W, 1982ApJ...253..386L, 1986ApJ...311.1015A, 1987ApJ...319L..27B, 1988ESASP.286..315T, 1988ApJ...335.1015B, 1992ApJ...393..782T, 1998ApJ...504.1029H}. The reduction is observed to depend on frequency, depth, and on the physical parameters that describe the flux tubes, i.e. their radius and magnetic strength \citep{1992ApJ...394L..65B, 1998ApJ...504.1029H, 2008ApJ...681..664G}. This behavior has been also found in small magnetic field concentrations \citep{1978SoPh...56....5R, 1981A&A....98..155S, 1996ApJ...465..406B, 1997ApJ...486L..67C, 2008ApJ...677..769H, 2009ApJ...695..325J, 2012ApJ...746...66H, 2012ApJ...744...98P}.\\ 
It is also known that at higher frequencies, above the acoustic cutoff frequency in the low photosphere, the velocity oscillations amplitude is enhanced at the edges of ARs \citep{1992ApJ...392..739B, 1992ApJ...394L..65B, 1993ApJ...415..847T}. Also, an enhancement of power in the three-minute band has recently been observed in the inner umbra of a pore \citep{2012A&A...539L...4S}.\\
Several mechanisms have been proposed to explain the observations \citep{1997ApJ...476..392H}:\\
1) The intrinsic power inhibition due to local convection suppression. By using numerical simulations \cite{2007ApJ...666L..53P} found that the reduction of wave excitation in a sunspot, can account up to $50\%$ of the power deficit.\\
2) Partial p-mode absorption \citep{1987ApJ...319L..27B, 1993ApJ...406..723B, 1995ApJ...451..372C, 2010SoPh..266...17C}. \cite{1992ApJ...391L.109S} showed that the wave absorption by sunspots can be interpreted in terms of p-modes mode conversion between the oscillations and MHD waves. Both models \citep{1993ApJ...402..721C} and simulations \citep{2008SoPh..251..291C} have shown that mode conversion is able to remove a significant amount of energy from the incident helioseismic wave, with an efficiency that depends on the angle of the magnetic field from the vertical, with a maximum at about $30^{\circ}$ \citep{2003MNRAS.346..381C, 2003SoPh..214..201C}.\\
3) Opacity effects. Within a magnetic region, the line-of-sight optical depth experiences a depression (Wilson depression). Due to increasing density, the amplitude of the velocity fluctuations decrease with depth.\\
4) Alteration of the p-mode eigenfunctions at the hands of the magnetic field \citep{1996ApJ...464..476J, 1997ApJ...476..392H}.

We note that the majority of the studies on the interaction between solar oscillations and magnetic field have focused on the use of models and simulations.\\
Because of the dependence of the velocity amplitude on magnetic strength, the most noticeable effects are observed within ARs, where the strongest magnetic fields are found. Among them, bipolar active regions (hereafter $\beta$ARs) are particularly interesting, showing asymmetries in their morphology and physical properties, as several works have already pointed out \citep[see, e.g., ][]{1979suns.book.....B, 1980A&A....92..111B, 1981SoPh...74..111T, 1985SoPh..100..397Z, 1990SoPh..126..285V, 1990SoPh..127...51P, 1993ApJ...405..390F}. \cite{1994SoPh..153..449M} conclude that morphological asymmetries are due to the different inclination with which the polarities of $\beta$ARs emerge.\\ 
Using simulations, \cite{1993ApJ...405..390F} have studied area asymmetries, and predict that the leading polarity must have a strength about twice larger than the trailing polarity, at the same depth. This result, which is a consequence of the Coriolis force action on the magnetic structure, is able to account for the trailing spot fragmentation, its lower flux magnitude and shorter lifetime.\\ 
MDI observations of 138 bipolar magnetic regions have recently quantitatively shown that the areas of leading polarities are typically smaller than those of trailing polarities \citep{2012A&A...539A..13Y}. This area asymmetry could be produced by the Coriolis force during a flux tube's rising motion in the solar convection zone.\\

In this work we investigated the velocity oscillation amplitude reduction by the magnetic field within $\beta$ARs. We analyzed 12 $\beta$ARs from SDO-HMI data, both in the Nothern (N) and Southern (S) hemispheres, and we found that the leading polarity systematically has a greater reduction in amplitude than the trailing. In addition, we studied the amplitude reduction as a function of the field inclination for AR11166. 

\section{SDO-HMI data}
\begin{figure}[ht!]
   \centering
   \subfigure[AR 11166]{\resizebox{\hsize}{!}{\includegraphics		{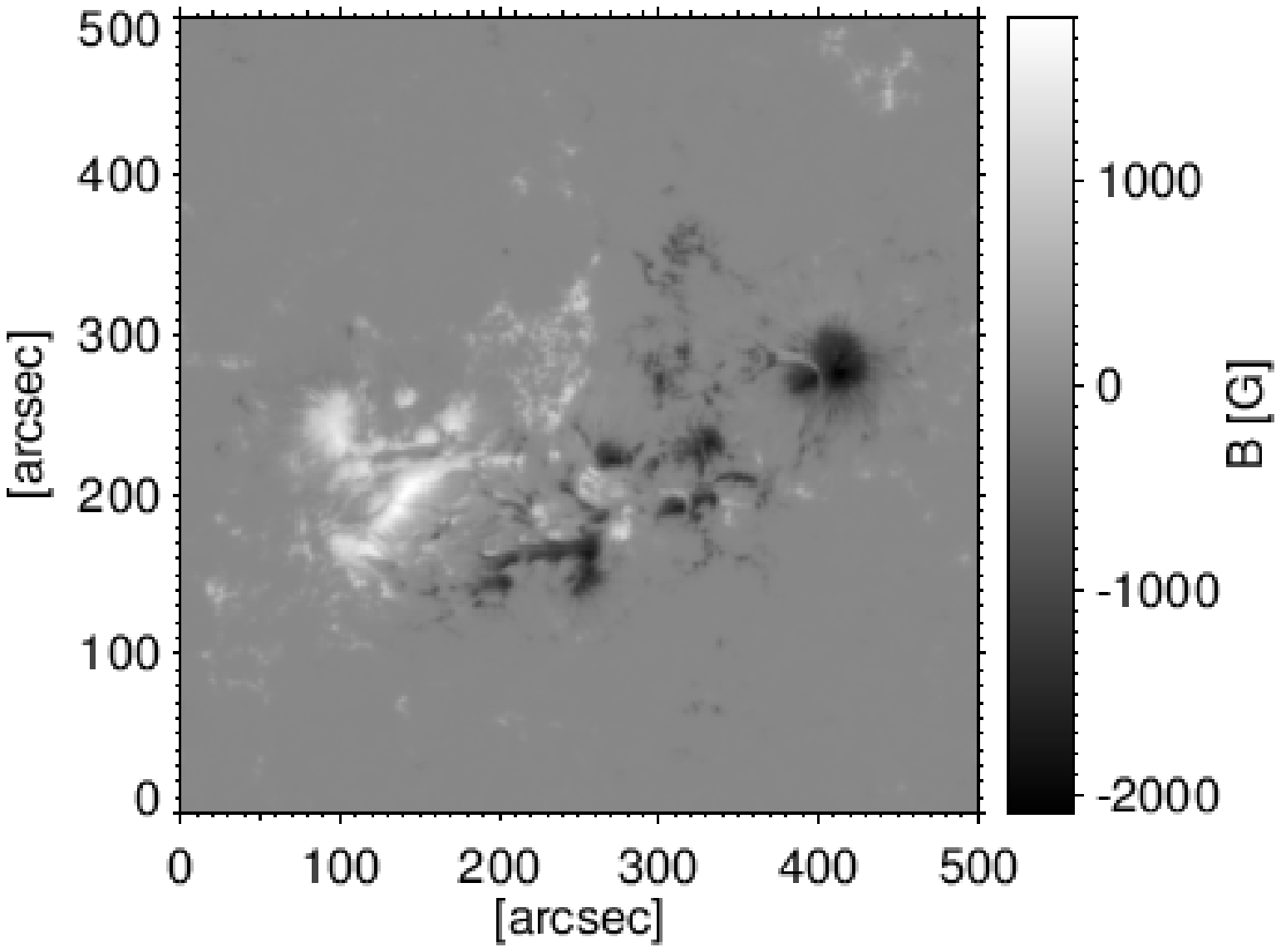}}}
   \subfigure[AR 11316]{\resizebox{\hsize}{!}{\includegraphics		{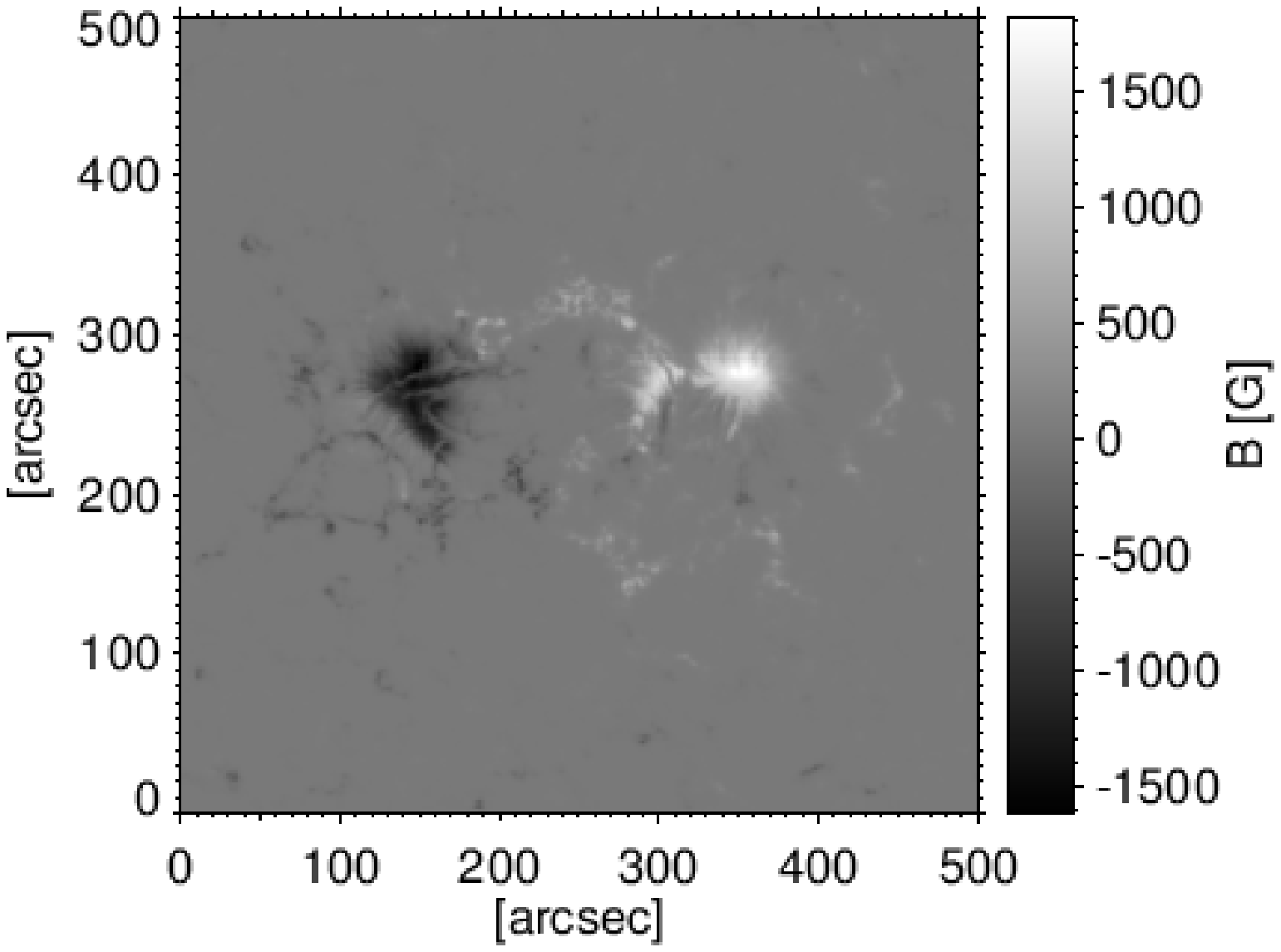}}}
   \caption{SDO-HMI mean magnetograms of (a) AR 11166 (b) AR 11316. See Table \ref{tab} for more details.} 
   \label{mags}
\end{figure}

Our data set consists of 12 SDO-HMI magnetograms and dopplergrams pairs with 1 arcsec spatial resolution \citep{springerlink:10.1007/s11207-011-9834-2, springerlink:10.1007/s11207-011-9842-2}. These were acquired with a 45 second cadence, which set the Nyquist frequency. Each $\beta$AR data set is 3 hours long, which set the lower cut-off at $\sim10^{-4} s^{-1}$. Observation times cover the interval from 2011 March 8 to 2012 January 3 (see Table 1).\\
We selected 8 isolated bipolar regions in the N hemisphere, and 4 in the S hemisphere. In order to limit the effects due to the inclination with respect to the line-of-sight, we selected $\beta$ARs as close as possible to the disk center. 
The rectangular area enclosing the $\beta$ARs was selected such that it included all the magnetic features with strength above 500 G, with the the minimum and maximum allowed dimensions of the enclosing rectangle being 500 and 700 arcseconds, respectively. 
Figure \ref{mags} shows the mean magnetograms of two selected $\beta$ARs: AR 11166 in the N hemisphere, AR 11316 in the S hemisphere. In both panels the leading polarity is on the right and it is negative for the Northern $\beta$ARs, positive for the Southern $\beta$ARs. Table 1 shows the list of $\beta$ARs analyzed in this work.
\begin{table}
\caption{Details of SDO-HMI data used. The fourth column shows the leading to trailing polarity area ratio.}
\label{tab}
\centering
  \begin{tabular}{ c c c c}
  \hline\hline
  $\beta$AR number & Observation Date & Location & $S_l/S_t$\\
  \hline
  \\
    AR11166 & 2011.03.08 & N11 W01 & 0.88\\
	    &				      & T11:00-14:00 &     \\
    AR11283 & 2011.09.05 & N13 W01 & 0.86\\
	    &		    & T17:00-20:00 &     \\
    AR11302 & 2011.09.28 & N13 W06 & 0.96\\
	    &				      & T17:30-20:30 &     \\
    AR11319 & 2011.10.15 & N09 E02 & 0.95\\
	    &		    & T07:00-10:00 &     \\
    AR11316 & 2011.10.15 & S12 E01 & 0.90\\
	    &		    & T07:00-10:00 &     \\
    AR11330 & 2011.10.27 & N09 E04 & 0.83\\
	    &		    & T18:30-21:30 &     \\
    AR11338 & 2011.11.06 & S14 W00 & 0.92\\
	    &		    & T20:00-23:00 &     \\
    AR11341 & 2011.11.11 & N08 W00 & 1.05\\
	    &		    & T07:00-10:00 &     \\
    AR11362 & 2011.12.03 & N08 W06 & 1.10\\
	    &		    & T20:00-23:00 &     \\
    AR11367 & 2011.12.07 & S18 W05 & 0.90\\
	    &		    & T20:00-23:00 &     \\
    AR11375 & 2011.12.13 & N08 W03 & 0.82\\
	    &		    & T19:00-22:00 &     \\
    AR11389 & 2012.01.03 & S22 W05 & 0.97\\
	    &		    & T16:00-19:00 &     \\
    \hline
  \end{tabular} 
\end{table}

\section{Method and analysis}
\label{meth}
The selected regions around each $\beta$AR were co-registered using a FFT technique ensuring sub-pixel accuracy. Inspection of the co-registered data shows that the strong magnetic structures remain in the same position during the entire duration of the acquisition.
The amplitude of velocity oscillations was estimated pixel-by-pixel through their integrated spectrum. Specifically, we selected a spectral window $\Delta\nu$ centered at $3$ mHz (five-minute band), having a width of $1$ mHz, and then we integrated the FFT amplitude spectrum $A_\nu$ in that frequency range
$$
A_{xy}=\int_{\Delta\nu}A_\nu d\nu
$$ 
We then divided the data into bins of magnetic field, each 25 G wide, and averaged the amplitude of velocity oscillations in each bin, thus obtaining $A_B=<A_{xy}>_{B}$. We consider only pixels with $|\mathbf{B}|>25$ G. Since all the amplitude distributions within the respective magnetic bin have a Gaussian-like shape (see inset in Figure \ref{pow}b), we use $\sigma/\sqrt{n}$ as error on $A_B$, where $\sigma$ is the standard deviation and $n$ are the counts in each bin.

\section{Results and Discussion}
\subsection{Oscillation amplitude VS magnetic field strength}
\label{powVSb}
\begin{figure}[t!]
   \centering
   \subfigure[AR 11166]{\resizebox{\hsize}{!}{\includegraphics		{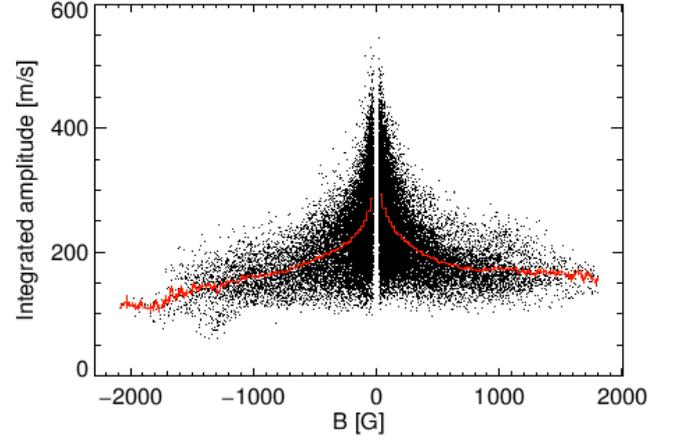}}}
   \subfigure[AR 11316]{\resizebox{\hsize}{!}{\includegraphics		{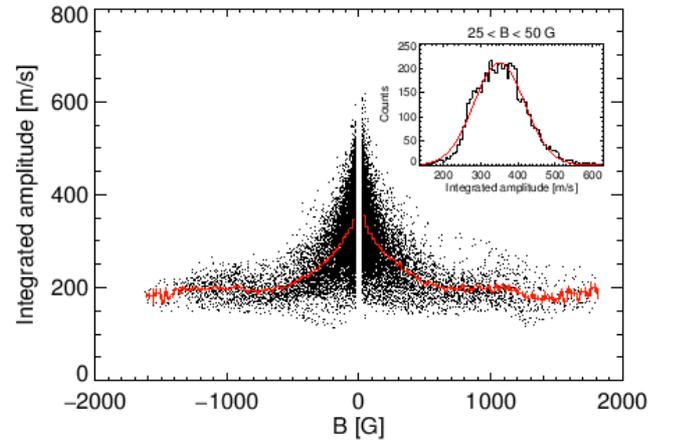}}}
   \caption{In panel a) the integrated velocity oscillation amplitude for AR11166 is shown; the same for AR1136 in panel b). The red overplots correspond to the average in each magnetic field bin. In the inset of panel b) amplitude distribution in the $25<B<50$ G bin (black line) and gaussan fit (red line) are shown.} 
   \label{pow}
\end{figure}

In Figure \ref{pow} we show the integrated amplitude $A_{xy}$ scatter plot (black dots), and $A_B$ for each magnetic bin (red line), for the $\beta$ARs shown in Figure \ref{mags}.  
Hereafter, we refer to the $A_B$ plot as the amplitude profile. The amplitude shows a decreasing trend up to a plateau, that is almost always reached in the range 0.5-1 kG (as in the case of Figure \ref{pow}b). This feature suggests that the amplitude is independent of the magnetic field strength above a threshold value, which is different for each $\beta$AR and each polarity in the same $\beta$AR.\\   
For every polarity present in the 12 $\beta$ARs analyzed, we computed the oscillation amplitude reduction as the ratio $d=A_{1000}/A_{25}$, where $A_{1000}$ and $A_{25}$ are, respectively, the mean oscillation velocity around $B=1000$ G, and that around $B=25$ G, the minimum observable strength.
Averaging the $d$ values over all the $\beta$ARs, we obtained $\bar{d}=0.54\pm0.06$ , which agrees with the oscillation amplitude reductions in magnetic environments quoted in the literature \citep[e.g.][]{1981SoPh...69..233W, 1982ApJ...253..386L, 1987ApJ...319L..27B, 1988ESASP.286..315T}.

The plateau in the amplitude profiles for high $B$ values, is intriguing. By visual inspection, we verified that it takes place almost exclusively in the $\beta$AR spot umbrae, as can be expected given the high magnetic field values. We can exclude an instrumental effect, since the magnetic strength threshold changes with each $\beta$AR and even with the polarity in the same $\beta$AR, and we are well above the 20 m/s error on the velocity measurements due to the HMI filter transmission profiles \citep{2011SoPh..271...27F}.  
It seems that the oscillation amplitudes are not completely suppressed even in the strongest magnetic fields, and are not dependent on the magnetic field strength anymore. Conversely to the simulations of \cite{2003ApJ...588.1183C}, it may indicate a saturation effect of the kinetic energy in the strong field regime, and shed new light onto the mechanisms that co-operate in the reduction of acoustic power in the lower atmosphere.\\

\subsection{The leading-trailing asymmetry}
\label{lt asym}
\begin{figure}[htbp]
 \centering
 \subfigure[]{\includegraphics		[width=8cm]{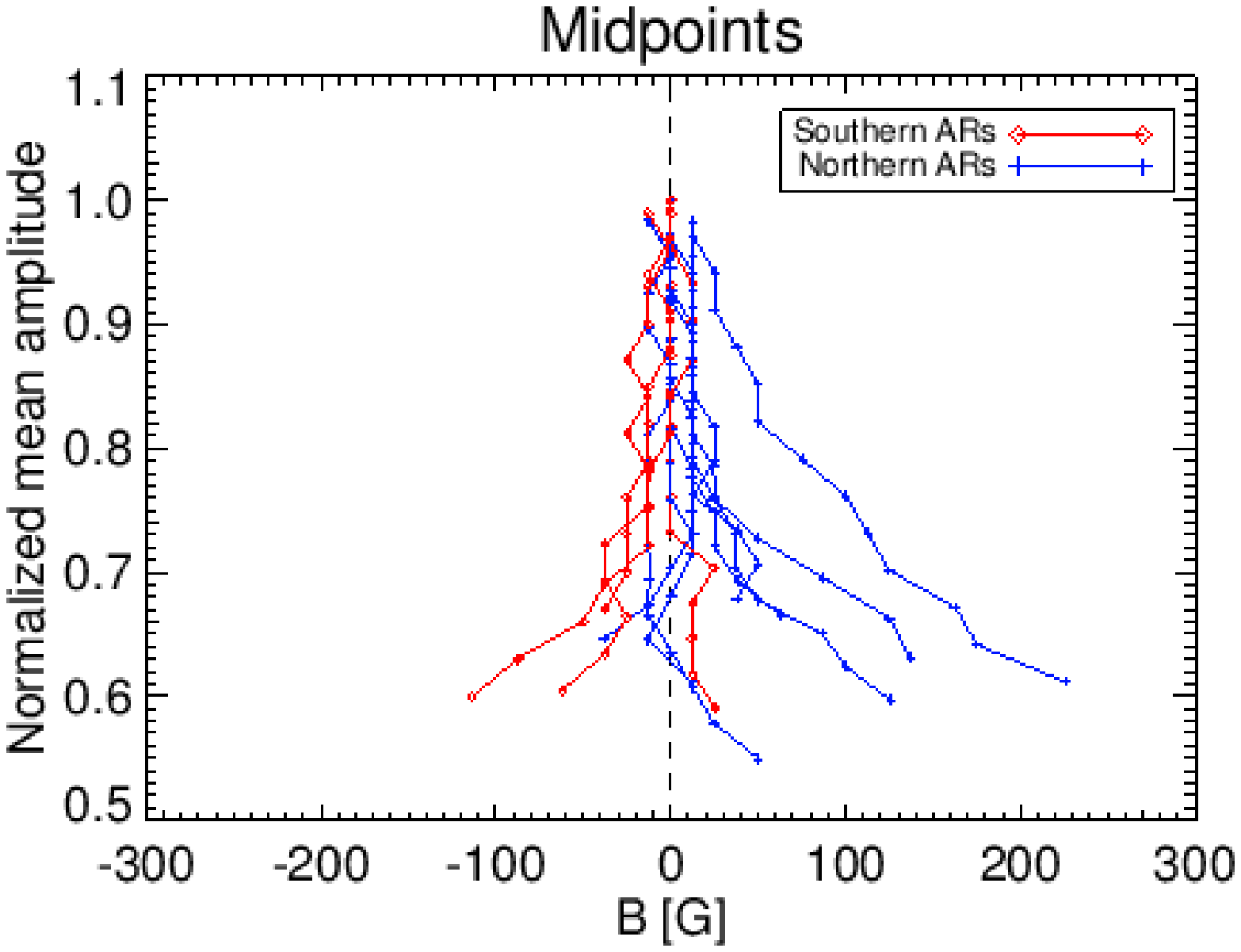}}
 \subfigure[]{\includegraphics		[width=8cm]{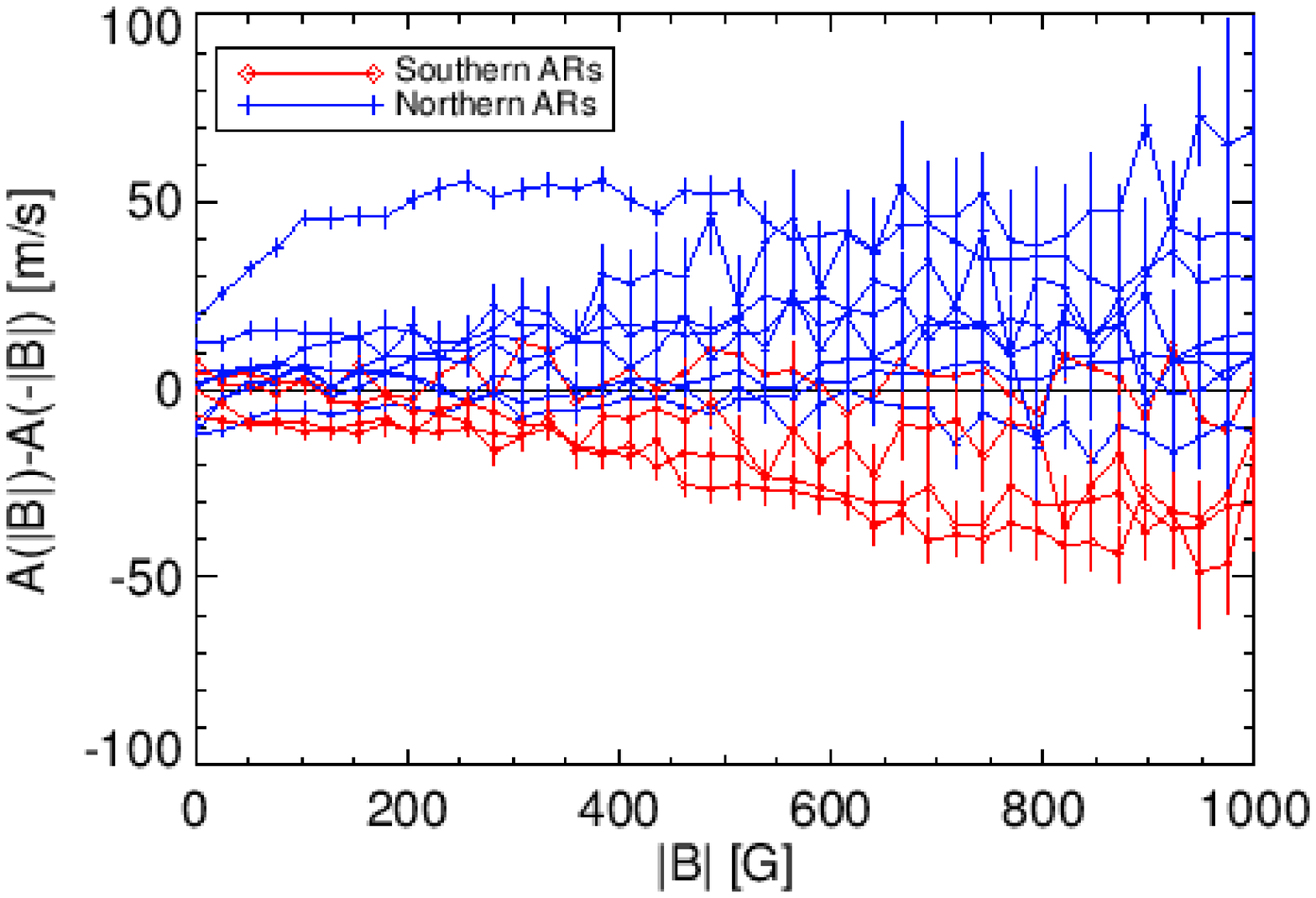}}
 \caption{a) Midpoints for normalized velocity oscillations amplitude. b) Difference between oscillation amplitude for both polarities against the magnetic strength. Bars represent errors. Red line with diamonds, refers to Southern $\beta$ARs, blue line with plus signs to Northern $\beta$ARs.}
 \label{bis}
\end{figure}

To investigate the relations between local oscillation amplitude and magnetic field strength, we compute the midpoints of the amplitude profiles. Specifically, we draw horizontal lines at constant $A_{B}$ values, proceeding down in steps of $10$ m/s.
We compute the abscissa of each midpoint on the segment that intersects the amplitude profiles.
If these profiles were symmetric, the midpoints should be on the $B=0$ vertical line.
In order to compare the amplitude profiles from different $\beta$ARs, we normalize them to their average value $A_{B}$ in the $B=\pm25$ G bins.
We remind that, for solar cycle 24, the leading polarities are negative in the N hemisphere and positive in the S hemisphere.
Figure \ref{bis}a shows the midpoint lines for all the 12 $\beta$ARs considered in this study. Northern and Southern $\beta$AR families appear well defined and separated. Moreover, as can be seen in Figure \ref{bis}, the amount of asymmetry, defined as the distance of each midpoint from the vertical line through $B=0$, increases for stronger fields. In the N hemisphere (in blue) it extends up to $|\mathbf{B}|\simeq230$ G, while in S hemisphere (in red) it is less apparent, and only reaches $|\mathbf{B}|\simeq130$ G. The amplitude profiles of Northern $\beta$ARs are biased toward positive polarities, i.e. the trailing, while the Southern $\beta$ARs are biased toward negative polarities, i.e. the trailing again.\\

To visualize differently the asymmetry, we can fix a value $|\mathbf{B}|$ for the magnetic strength, and compute the amplitude difference, i.e. the difference between the respective averaged amplitudes for both polarities in each $\beta$AR, namely $A_{|\mathbf{B}|}-A_{-|\mathbf{B}|}$. In Figure \ref{bis}b we show these differences versus the absolute value of magnetic strength for all $\beta$ARs of Table \ref{tab}. 
This plot confirms that the asymmetry is lower where the field is weaker. At $25$ G the difference is about $10$ m/s and typical values for velocity amplitudes up to $600$ m/s; at $1000$ G the difference rises to over $50$ m/s, while the amplitudes are $\sim200$ m/s. 
Both in N and S hemispheres the oscillation amplitude in the trailing polarity is higher than that in the leading polarity. This fact implies that $A_{|\mathbf{B}|}>A_{-|\mathbf{B}|}$ in the N hemisphere and viceversa in the S hemisphere, and that the amplitude reduction is less efficient in the trailing polarity.

Such an asymmetry suggests that the amount of oscillation reduction within magnetic environments is not a function of the local magnetic field strength only.
If this were the case, we would observe the same oscillation amplitudes for $\mathbf{B}$ and $\mathbf{-B}$ magnetic fields. This would imply both the plots of Figure \ref{bis} to lie on the line through $0$ G (a) and $0$ m/s (b), respectively.\\

The simulations of \cite{1993ApJ...405..390F} about area asymmetries of $\beta$ARs, predicted a pronounced difference in the field strength between the leading and the trailing polarities, and ensuing differences in their fragmentation and lifetimes.
This suggests that the asymmetry we found could derive from non-local properties, i.e. on the whole magnetic environment topology and even its surroundings.\\
We therefore computed the area asymmetry in our magnetograms following \citet{2012A&A...539A..13Y}, with a threshold $B>500$ G. The results are reported in the fourth column of Table \ref{tab}.
We found that, on average, the area of the leading polarity is $\sim$90\% of the trailing polarity area, regardless the hemisphere which the $\beta$AR belongs to. 
This quantifies how much a trailing polarity is spread with respect to its leading polarity.
We can speculate that a relation may exist between the area and the oscillation suppression, due to a different interaction of acoustic waves with sparser or denser active region polarities.\\ 

\subsection{Oscillation amplitude VS magnetic field inclination\\ The case of AR11166}
\label{powVSincl}
To support the findings reported in sec. \ref{lt asym}, we studied the dependence of the oscillation amplitude on the magnetic field inclination angle $\theta$ in an $\beta$AR. For this purpose, we considered the AR11166 (Figure 1a) as a case study. 
We used the LOS inclination maps provided by the VFISV \citep[Very Fast Inversion of the Stokes Vector,][]{springerlink:10.1007/s11207-010-9515-6} inversion of HMI vector magnetic field data. \\
\begin{figure}[ht!]
   \centering
   \resizebox{\hsize}{!}{\includegraphics		{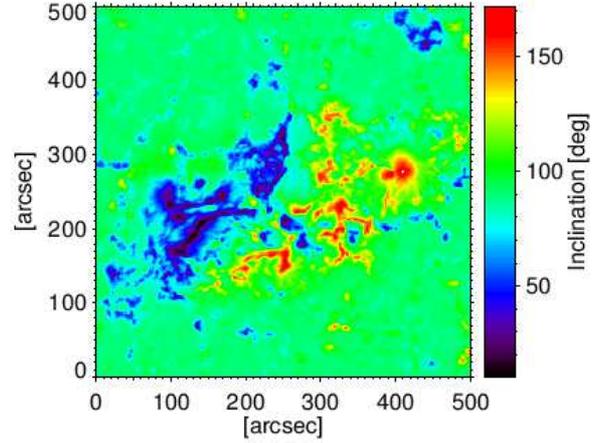}}
   \caption{Inclination map of AR 11166. The leading polarity (negative) ranges from $90^\circ$ to $180^\circ$, and the trailing polarity (positive) from $0^\circ$ to $90^\circ$.} 
   \label{incmap}
\end{figure}
In Figure \ref{incmap} we show the magnetic field inclination map of AR11166. The inclinations retrieved by the VFISV span the range $10^\circ < \theta < 172^\circ$. Following the most used sign convention, the leading polarity (negative) ranges from $90^\circ$ to $180^\circ$, and the trailing polarity (positive) from $0^\circ$ to $90^\circ$. As expected, the innermost umbrae are mostly line of sight, as the $\beta$ARs were selected as close as possible to the disk center.\\
VFISV also provides the error $\sigma_\theta$ associated to the inclination, for each pixel \citep[][]{springerlink:10.1007/s11207-010-9515-6, 1986nras.book.....P}.
To avoid polarity flips due to the inversion error, we discarded all those pixels where $\theta\pm3\sigma_\theta$ causes the field to flip its inclination from positive to negative (or vice-versa).
We stress that VFISV is a Milne-Eddington code, therefore blind to any $B$ and $v$ line of sight gradients, which produce asymmetries in the Stokes profiles. This happens most often in regions filled by weak fields \citep{2011A&A...526A..60V, 2011A&A...530A..14V}.
Moreover, the inversion of low signal to noise Stokes profiles is usually problematic \citep[see ][for more details]{1992soti.book...71L, springerlink:10.1007/s11207-010-9515-6}. 
For noisy Stokes profiles, and therefore low B values, VFISV (and any inversion code) tends to be biased toward $\sim90^{\circ}$ inclinations \citep{2011A&A...527A..29B}. For these reasons and to be consistent with the threshold used in \ref{meth}, we rejected the pixels with $B<25$ G and $75^\circ < \theta < 105^\circ$. After this selection, about $17\%$ of the initial pixels were selected for the analysis.

\begin{figure}[ht!]
 \centering
\includegraphics		[width=8cm]{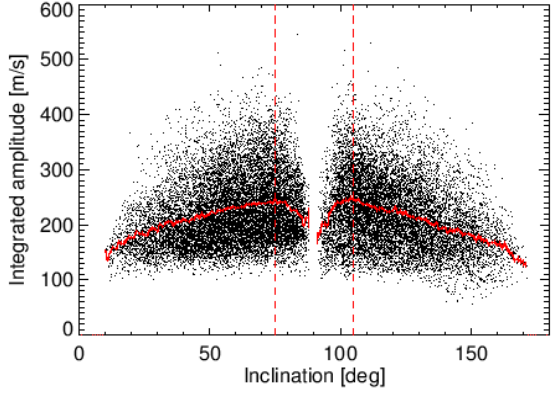}
\caption{Integrated velocity oscillation amplitude within AR11166 as a function of the magnetic field inclination $\theta$. The thick red line corresponds to the average amplitude $A_\theta=<A_{xy}>_\theta$ in each $\theta$ bin ($\Delta\theta=1^\circ$). The vertical dashed lines mark the threshold $\theta < 75^\circ$ OR $\theta>105^\circ$.}
 \label{pow_vs_inc}
\end{figure}

\begin{figure}[ht!]
 \centering
\includegraphics		[width=8cm]{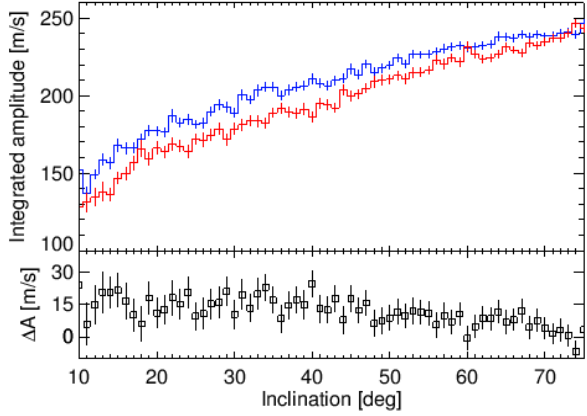}
\caption{{\itshape Upper panel:} $A_\theta^T$ and $A_\theta^L$ versus $\theta$ in the range $[10^\circ,75^\circ]$. 
The blue line represents the average amplitude of the positive (trailing) polarity: $A_\theta^T$. The red line represents the average amplitude of the negative (leading) polarity after a remapping operation: $\theta\rightarrow 180^\circ-\theta$: $A_\theta^L$. {\itshape Lower panel:} $\Delta A$ versus $\theta$ (see the text). Vertical bars represent the error, which were computed for each $A_\theta$ with the same method described in section \ref{meth}, and then summed up.}
 \label{asym_incl}
\end{figure}

We plotted $A_{xy}$ and its average in each $\theta$ bin $A_\theta=<A_{xy}>_\theta$ against the magnetic field inclination $\theta$ (shown in Figure \ref{pow_vs_inc}). 
For ease of comparison, in the upper panel of Figure \ref{asym_incl} we show again $A_\theta$ versus $\theta$ in the range $[10^\circ,75^\circ]$. 
Both $A_\theta$ smoothly decrease from $\simeq250$ m/s to $\sim140$ m/s, from almost horizontal to almost vertical fields, respectively, but the oscillation amplitude reduction is more effective for the leading polarity (red curve) than for the trailing polarity (blue curve).\\
In the lower panel of Figure \ref{asym_incl} we plot the oscillation amplitude asymmetry $\Delta A=A_\theta^T-A_\theta^L$ VS $\theta$.
$\Delta A$ is constant ($\simeq15$ m/s) within the errors up to $\theta\simeq45^\circ$, then $\Delta A$ drops to very few m/s.\\
The inclination analysis shows that a $\simeq15$ m/s asymmetry exists between the polarities of AR11166. In particular, the velocity oscillation amplitude is enhanced in the trailing polarity with respect to the leading.\\

There is no evident relation between the asymmetry $\Delta A$ and the inclination $\theta$. For any $\theta\lesssim45^\circ$ the oscillation amplitude asymmetry is positive and around $15$ m/s.
We interpret the drop at $\simeq45^\circ$ as the combined effect of the VFISV preference to associate weak fields with larger inclinations \citep{2011A&A...527A..29B}, and of the amplitude difference to be small for weak fields (see Figure \ref{bis}b).
Also, an incorrectly retrieved weak $\vec{B}$ may result due to unresolved magnetic structuring in the pixel \citep[e.g. ][]{1998ASPC..155...54S, 2003ApJ...593..581S, 2011A&A...526A..60V}, and we recall that a VFISV hypothesis is that the magnetic field is constant within the pixel.\\
\citet{2006MNRAS.372..551S} and \citet{2011A&A...534A..65S} have demonstrated that, due to mode conversion, there exists a preferred angle at which the power is significant larger. In our case, we indeed focus on the amplitude asymmetry between the two polarities, instead of considering the proper amplitude dependence that may be affected by biases due to the magnetic field strength.

\section{Conclusions} 
In this work we focused on the velocity oscillation amplitude reduction in $\beta$ARs, reaching the following conclusions.\\

\noindent 
The oscillation amplitude reduction found in magnetic environments of $\beta$ARs is $0.54\pm0.06$, which agrees with the values quoted in the literature \citep[e.g.][]{1981SoPh...69..233W, 1982ApJ...253..386L, 1987ApJ...319L..27B, 1988ESASP.286..315T}.\\

\noindent  
The five-minute amplitudes are not completely suppressed in the strongest magnetic fields of the spot innermost umbra. It would be very important to understand why a plateau appears in the oscillation amplitude profiles for strong fields.\\

\noindent   
There exists a leading-trailing polarity asymmetry in $\beta$ARs. The asymmetry suggests that the reduction in oscillation amplitude does not depend on the field strength only, but may depend also on non-local conditions, such as the area on which the field spreads, for instance.\\

\noindent  
The trailing polarity systematically shows a higher oscillation amplitude than the leading polarity, regardless the hemisphere which the $\beta$AR belongs to.\\

\noindent  
The plot of the velocity oscillation amplitude as a function of the magnetic field inclination confirms such an asymmetry in $\beta$ARs. Apparently, the asymmetry does not evidently depend on the inclination.\\

We took advantage from HMI full-disk data at a $45$ s time cadence, and at $1$ arcsec spatial resolution. The analysis of these data revealed a possible saturation of the oscillation amplitude reduction for strong $B$, and an asymmetry in such a reduction for leading-trailing polarities. These results have to be accounted for in modeling the power reduction in magnetic environments, the emergence and the evolution of $\beta$ARs.

\begin{acknowledgements}
  We thank professor Stuart M. Jefferies (IfA, University of Hawaii) for useful discussion and critical reading of the early version of this manuscript.\\ 
  SDO-HMI data are courtesy of the NASA/SDO HMI science team. We acknowledge the VSO project (http://vso.nso.edu) through which data were easily obtained.  
\end{acknowledgements}
\bibliography{giannattasio}
\bibliographystyle{aa}
\end{document}